\documentclass[journal]{IEEEtran}

\ifCLASSINFOpdf
\else
\fi

\usepackage{mathrsfs}
\usepackage{pifont}
\usepackage{bbding}
\usepackage{tipa}
\usepackage{color}
\usepackage{epsfig}
\usepackage{graphicx}
\usepackage{url}
\usepackage{amsfonts}
\usepackage{multicol}
\usepackage{float}
\usepackage{times}
\usepackage{psfrag}
\usepackage{subfigure}
\usepackage{stfloats}
\usepackage{amsmath}
\usepackage{footnote}
\usepackage{array}
\usepackage{amsmath,epsfig}
\usepackage{stmaryrd}
\usepackage{amssymb}
\usepackage{epic}
\usepackage{graphicx}
\usepackage{curves}
\usepackage{balance}
\usepackage{graphicx}
\usepackage{epstopdf}
\usepackage{booktabs}
\usepackage{algpseudocode}
\usepackage{algorithm}
\usepackage{caption}
\usepackage{cite}
\usepackage{caption}
\usepackage{bm}

\begin{document}

\captionsetup[figure]{name={Fig.},labelsep=period}

\title{Path Planning for UAV-Mounted Mobile Edge Computing with Deep Reinforcement Learning}
\author{Qian Liu, Long Shi, Linlin Sun, Jun Li,  Ming Ding, and Feng Shu

\thanks{ Q. Liu, L. Sun, J. Li, and F. Shu are with School of Electronic and Optical Engineering, Nanjing University of Science and Technology, Nanjing, China. E-mail:\{qianliu6767, sunlinlin, jun.li, shufeng\}@njust.edu.cn. L. Shi is with Science and Math Cluster, Singapore University of Technology and Design, Singapore. E-mail:slong1007@gmail.com. M. Ding is with Cyber-Physical Systems, Data61, Sydney, Australia. E-mail: ming.ding@data61.csiro.au.}}
\maketitle

\begin{abstract}
In this letter, we study an unmanned aerial vehicle (UAV)-mounted mobile edge computing network, where the UAV executes computational tasks offloaded from mobile terminal users (TUs) and the motion of each TU follows a Gauss-Markov random model. To ensure the quality-of-service (QoS) of each TU, the UAV with limited energy dynamically plans its trajectory according to the locations of mobile TUs. Towards this end, we formulate the problem as a Markov decision process, wherein the UAV trajectory and UAV-TU association are modeled as the parameters to be optimized. To maximize the system reward and meet the QoS constraint, we develop a QoS-based action selection policy in the proposed algorithm based on  double deep Q-network. Simulations show that the proposed algorithm converges more quickly and achieves a higher sum throughput than conventional algorithms.
\end{abstract}

\begin{IEEEkeywords}
Unmanned aerial vehicle, edge computing, path planning, Markov decision process, deep reinforcement learning
\end{IEEEkeywords}

\IEEEpeerreviewmaketitle

\section{Introduction}
Mobile edge computing (MEC) enables the computational power at the edge of cellular networks to flexibly and rapidly deploy innovative applications and services towards mobile terminal users (TUs) \cite{MEC_survey}.
In contrast to position-fixed edge servers, recent works on MEC have been devoted to mobile edge servers that can provide more flexible and cost-efficient computing services in hostile environments.
As a moving cloudlet, the unmanned aerial vehicle (UAV) can be applied in MEC due to its reliable connectivity with affordable infrastructure investment \cite{UAV-MEC-survey}.
For example, \cite{3-1} proposed an adaptive  UAV-mounted cloudlet-aided recommendation system in the location based social networks to provide active recommendation for mobile users.
Recently, \cite{3-3} proposed a distributed anticoordination game based partially overlapping channel assignment algorithm in the UAV-aided device-to-device networks to achieve good throughput and low signaling overhead.
Later on, \cite{3-2} developed a novel game-theoretic and reinforcement learning (RL) framework in the UAV-enabled MEC networks, in order to maximize each base station's long-term payoff by selecting a coalition and deciding its action.

Recent research mainly focuses on path planning in the UAV-mounted MEC networks.
For instance, \cite{UAV-mounted-MEC_path-planning} jointly optimized the UAV trajectory and bit allocation under latency and UAV energy constraints.
Later on, \cite{UAV-assited-MEC_fixed-trajectory} studied a fixed UAV trajectory with  dynamic power allocation  among the social internet of vehicles.
On one hand, the UAV trajectories were designed offline in \cite{UAV-mounted-MEC_path-planning,UAV-assited-MEC_fixed-trajectory,WFF}, assuming that the TU locations are invariant. However, the TU locations may change dynamically over time in practice.
To ensure the quality-of-service (QoS) for each TU, the UAV needs to adjust its trajectory according to the time-varying TU locations.
How to design the UAV trajectory to serve mobile TUs in the MEC networks remains challenging and primarily motivates our work.
On the other hand, the trajectory optimization relies on either dynamic programming \cite{UAV-mounted-MEC_path-planning} or successive convex approximation method \cite{UAV-assited-MEC_fixed-trajectory}\cite{WFF}.
A major concern lies in that the optimization for the offline trajectory designs in \cite{UAV-mounted-MEC_path-planning,UAV-assited-MEC_fixed-trajectory,WFF} may not be feasible to deal with the mobile TUs in MEC networks.

Markov decision process (MDP) and RL algorithm have been applied in online UAV trajectory design to improve the detection accuracy \cite{MDP_fuzzy_Q_PP} and detect locations of endangered species \cite{IOT-QL-path-planning}.
However, the dynamic change of TU locations inevitably leads to innumerable states in the MDP, making the path planning problem even more complex.
In this context, deep reinforcement learning (DRL) algorithm is more adequate to deal with the curse of huge state and action spaces induced by time-varying TU locations   than conventional RL methods.
 Ref. \cite{MCS} leveraged DRL for enabling model-free UAV control to collect the data from users in mobile crowd sensing-based smart cities.
Recently, \cite{Space/Aerial-Assisted+MEC} investigated a joint resource allocation and task scheduling approach in a space-air-ground integrated network based on policy gradient and actor-critic methods, where the UAVs provide near-user edge computing for static TUs.
Moreover, \cite{UAV-Aided-Communications} proposed the deterministic policy gradient algorithm to maximize the expected uplink sum rate in the UAV-aided cellular networks with mobile TUs.
Among the value based DRL algorithms,
\cite{DDQN} unveiled that double deep Q-network (DDQN) addresses the overestimation problem in deep Q-network (DQN) via decoupling target Q-value and predicted Q-value, and generates a more accurate state-action value function than DQN.
It is known that the better state-action value function corresponds to the better policy.
Under this policy, the agent chooses the better action to improve the system reward.

In this letter, we propose a DRL-based algorithm for the UAV to serve the mobile TUs in the UAV-mounted MEC network, where the motion of each TU follows the Gauss-Markov random model (GMRM).
Our goal is to optimize the UAV trajectory to maximize the long-term system reward subject to limited energy of UAV and QoS constraint of each TU.
Toward this goal, we formulate the optimization problem as an MDP.
In particular, we develop a QoS-based $\epsilon$-greedy policy in our proposed algorithm to maximize the system reward and meet the QoS constraint.
Simulation results show that our proposed algorithm outperforms conventional RL and DQN algorithms in terms of convergence and throughput, and the QoS-based $\epsilon$-greedy policy can achieve $99\%$ guarantee rate in QoS of each TU.

\section{System Model}
Fig. 1 shows that a UAV with limited energy $B$ provides computational services to $N$ TUs over a certain period.
The operating period is discretized into $T$ times slots each with non-uniform duration, indexed by $t=0,1,2...,T-1$.
Suppose that the UAV can only serve a single TU in each time slot, referred to as the association between the UAV and TU.
In each time slot, the UAV can only hover over one of $M$ fixed perceptual access points (FPAPs) to form direct connection with the associated TU and execute its offloaded tasks.
\begin{figure}[t]
\centering
\includegraphics[width=2.8 in]{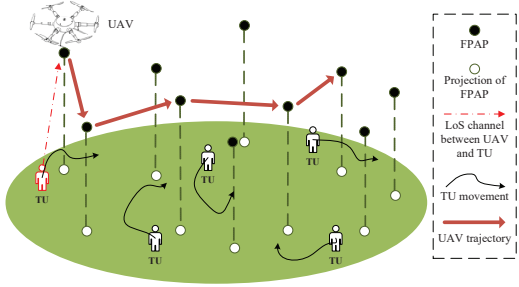}
\captionsetup{font={small}}                   
\caption{The UAV-mounted MEC network, where the UAV hovers over $M$ FPAPs to serve the $N$ mobile TUs.}
\label{fig:system_model}
\end{figure}

\subsection{Movement Model of TUs}
Consider that all TUs are randomly located at $t=0$.
Assume that all TU locations do not change during the duration $\Delta_{t,t-1}$ between the $t$th and $t-1$th time slots.
Following the GMRM in \cite{GMRM}, the velocity $v_n(t)$ and direction $\theta_n(t)$ of the $n$th TU in the $t$th time slot ($t\geq 1$) are updated as
\begin{subequations}
\begin{equation}
\begin{array}{l}
{v_n(t)} = \kappa_1 {v_n(t-1)} + (1-\kappa_1 ) \bar{v}  + \sqrt {1-{\kappa_1 ^2}}  {\Phi_n}, \tag{1a}
\end{array}
\end{equation}
\begin{equation}
\begin{array}{l}
{\theta_n(t)} = \kappa_2 {\theta_n(t-1)} + (1-\kappa_2 ) \bar{\theta}_n  + \sqrt {1 - {\kappa_2 ^2}}  {\Psi_n}, \tag{1b}
\end{array}
\end{equation}
\end{subequations}
where $0\leq \kappa_1, \kappa_2\leq 1$ are utilized to adjust the effect of the previous state, $\bar{v}$ is the average velocity for all TUs, and $\bar{\theta}_n$ is the average direction of the $n$th TU.
In particular, we consider that the average speed for all TUs is same and different TUs have distinct average directions.
Also, $\Phi_n$ and $\Psi_n$ follow two independent Gaussian distributions with different mean-variance pairs $(\bar{\xi}_{v_n}, \varsigma_{v_n}^2)$ and $(\bar{\xi}_{\theta_n}, \varsigma_{\theta_n}^2)$ for the $n$th TU, both of which reflect the randomness in the movements of different TUs.
Let $l_n^{{\rm{TU}}}(t) = [{x_n^{{\rm{TU}}}}(t),{y_n^{{\rm{TU}}}}(t)]$ denote the location of the $n$th TU in the $t$th time slot. Given (1a) and (1b), the TU location is updated as \cite{GMRM}
\begin{subequations}
\begin{equation}
x_n^{{\rm{TU}}}(t) = x_n^{{\rm{TU}}}(t-1) + {v_n}(t-1)\cos ({\theta _n}(t-1))\Delta_{t,t-1}, \tag{2a}
\end{equation}
\begin{equation}
y_n^{{\rm{TU}}}(t) = y_n^{{\rm{TU}}}(t-1) + {v_n}(t-1)\sin ({\theta _n}(t-1))\Delta_{t,t-1}. \tag{2b}
\end{equation}
\end{subequations}
Also, the UAV location at the $m$th FPAP  in the $t$th time slot is $l_m^{{\rm{UAV}}}(t) = [x_m^{{\rm{UAV}}}(t),y_m^{{\rm{UAV}}}(t)],{m} \in {\{ 1,2,...,M\}}.\label{UAV_location}$

\subsection{Energy Consumption of UAV}
The energy consumption of the UAV falls into the following three categories:

(1) {\emph{Flying Energy Consumption $e_\emph{f}(t)$}:}
Let $V$ and $P_\text f$ denote the UAV flying speed and the UAV flying power respectively.
Consider that $V$ is constant over $T$ time slots. Moreover, $P_\text f=P_\text P + P_\text I$, where $P_\text P$ and $P_\text I$ denote the parasitic power and the induced power to overcome the parasitic drag and the lift-induced drag respectively\cite{Flying_model}.
Consequently, the flying energy consumed by the UAV flying from one FPAP in the $t-1$th time slot to another in the $t$th time slot is given by
\begin{equation}
\begin{array}{l}
e_{\text f}(t) = {P_\text f}\frac{{\sqrt {{{[{x_m^{{\rm{UAV}}}}(t)- {x_m^{{\rm{UAV}}}}(t-1)]}^2} + {{[{y_m^{{\rm{UAV}}}}(t) - {y_m^{{\rm{UAV}}}}(t-1)]}^2}}}}{V}.
\end{array}
\end{equation}

(2) {\emph{Hovering Energy Consumption $e_\emph{h}(t)$}:}
Considering the line-of-sight channel between the UAV and its associated TU, the uploading rate (bits/s/Hz) from the associated $n$th TU to the UAV at the $m$th FPAP in the $t$th time slot is given by
\begin{equation}
\begin{array}{l}
{R_{nm}}(t) = {\log _2}\left(1 + \frac{{{P_\text t}{c_{nm}}(t)}}{{{\sigma ^2}}}\right),\label{rate}
\end{array}
\end{equation}
where $P_\text t$ is the transmission power of each TU, $\sigma^2$ is Gaussian white noise power at the UAV, and ${c_{nm}(t)} = \frac {\rho_{0}}{\sqrt {{H^2} + {{[{x_m^{{\rm{UAV}}}}(t) - {x_n^{{\rm{TU}}}}(t)]}^2} + {{[{y_m^{{\rm{UAV}}}}(t) - {y_n^{{\rm{TU}}}}(t)]}^2}}}$ denotes the channel gain between the $n$th TU and the $m$th FPAP with $\rho_{0}$ being the path loss per meter and $H$ being the fixed flying altitude of the UAV.
From (\ref{rate}), the hovering energy consumed by the UAV in the $t$th time slot is given by
\begin{equation}
{e_\text h}(t) = {P_\text h}\frac{{{\mu _n}(t)}{N_\text b}}{{{R_{nm}}(t)}},
\end{equation}
where $P_\text h$ is the UAV hovering power, $\mu _n(t)$ is the amount of offloaded tasks from the $n$th TU in the $t$th time slot, and $N_{\text{b}}$ is the number of bits per task.

(3) {\emph{Computing Energy Consumption $e_\emph{c}(t)$}:}
The computing energy for the offloaded tasks from the $n$th TU is ${e_\text c}(t) = \gamma_\text c {C}{({f_\text c})^2}{\mu _n}(t){N_\text b},$ where $\gamma_\text c$ is the effective switched capacitance, $C$ is the number of CPU cycles for computing one bit, and $f_{\text c}$ is the CPU frequency \cite{WQQ}.

Consequently, the total energy consumption of the UAV in the $t$th time slot is $W(t) = {e_{\rm{f}}}(t) + {e_{\rm{h}}}(t) + {e_{\rm{c}}}(t)$, and the energy that can be used by the UAV in the $t+1$th time slot is
\begin{equation}
\begin{array}{l}
b(t+1) = b(t) - W(t).\label{battery}
\end{array}
\end{equation}

\section{MDP Modeling and Problem Formulation}
From (2a), (2b) and (\ref{battery}), the locations of TUs and the UAV energy possess Markov characteristics. As such, we formulate the optimization problem of the UAV trajectory as an MDP. Our goal is to maximize the long-term system reward subject to the UAV energy and TUs' QoS constraint.

\subsection{State, Action, and Reward}
The state space of MDP is described as
\begin{equation}
\begin{array}{l}
\mathcal{S}=\Big \{ s_t|s_t =\{ l_n^{{\rm{TU}}}(t), l_m^{{\rm{UAV}}}(t), c_{nm}(t), b(t)\}, \\
{n} \in {\{ 1,2,...,N\}}, {m} \in {\{ 1,2,...,M\}}, {t} \in {\{ 0,1,...,T-1\}} \Big \}.
\end{array}
\end{equation}
Furthermore, the UAV chooses to serve one of $N$ TUs among one of $M$ FPAPs in each time slot.
Overall, the action space in our system includes two kinds of actions, denoted by
\begin{equation}
\begin{array}{l}
\mathcal{A} = \Big \{a_{nm}(t)|a_{nm}(t)=\{a_{n}^{\text{TU}}(t), a_{m}^{\text{FPAP}}(t)\}, \\
{n} \in {\{ 1,2,...,N\}}, {m} \in {\{ 1,2,...,M\}}, {t} \in {\{ 0,1,...,T-1\}}\Big \},
\end{array}
\end{equation}
where $a_{n}^{\text{TU}}(t)$ represents that the UAV chooses the $n$th TU in the $t$th time slot and $a_{m}^{\text{FPAP}}(t)$ represents that the UAV flies to the $m$th FPAP in the $t$th time slot.

Suppose that the UAV serves the $n$th TU in the $t$th time slot.
In general, system utility is closely related to the number of offloaded tasks $\mu_n(t)$.
However, the correlation is not simply in a linear manner.
With reference to \cite{non_convex_utility}, we adopt a sigmoidal-like function to describe the correlation as
\begin{equation}
U({\mu _n}(t)) = 1 - \exp \left[ { - \frac{{{{({\mu _n}(t))}^\eta }}}{{{{{\mu _n}(t)}} + \beta }}} \right],\label{utility}
\end{equation}
where the constants $\eta$ and $\beta$ are used to adjust the efficiency of $U({\mu _n}(t))$.
Note that the values of $\eta$ and $\beta$ vary as the range of ${\mu _n}(t)$ changes.
From (\ref{utility}), the system utility first increases steeply as ${\mu _n}(t)$ rises and then becomes steady when ${\mu _n}(t)$ is sufficiently large. Therefore, the heuristic use of (\ref{utility}) prevents the UAV from serving any single TU over a long period while ignoring other TUs, which is consistent with the QoS constraint in (\ref{constraint}).
In addition, the system reward takes the effect of UAV energy consumption into account.
As such, the system reward in the $t$th time slot induced by the current state $s_t$ and action $a_t$ is defined as
\begin{equation}
r_{t+1} = U({\mu _n}(t)) - \psi W(t),\label{reward}
\end{equation}
where $\psi {\rm{ = }}\frac{{\rm{1}}}{{\max \limits_{t} {\kern 1pt} W(t)}}$ is used to normalize $W(t)$ and unify the unit of $W(t)$ and $U({\mu _n}(t))$.

\subsection{Problem Formulation}

From \cite{DDQN}, the policy in RL corresponds to the probability of choosing the action $a_t$ according to the current state $s_t$.
The optimal policy $\pi^*$ is the specific policy that contributes to the maximal long-term system reward.
Our goal is to find $\pi^*$ to maximize the average long-term system reward as
\begin{subequations}
\begin{equation}
\mathop {{\rm{arg}{\kern 1pt} \rm{max}}{\kern 1pt} {\kern 1pt} {\kern 1pt} {\kern 1pt} {\kern 1pt} }\limits_{{\pi ^*}} \frac{{\sum\nolimits_{t = 0}^{T-1} {{r_{t + 1}}} }}{T} \tag{11a}
\end{equation}
\vspace{-0.2cm}
\begin{equation}
{\rm{s}}{\rm{.t}}{\rm{.}}\;\;  \sum\nolimits_{t = 0}^{T-1} {W(t)}  \le B, {\kern 1pt} {\kern 1pt} \sum\nolimits_{t = 0}^{T-1} {{\mu _n}(t)} \ge Z,\; \forall n, \label{constraint} \tag{11b}
\end{equation}
\end{subequations}
where the first constraint represents that the total energy consumption over $T$ time slots cannot exceed the UAV battery capacity and the second constraint (i.e., QoS constraint) guarantees the minimum amount of offloaded tasks (i.e., $Z$) from each TU over $T$ time slots.

\begin{figure}[ht]
\centering
\includegraphics[width=2.8in]{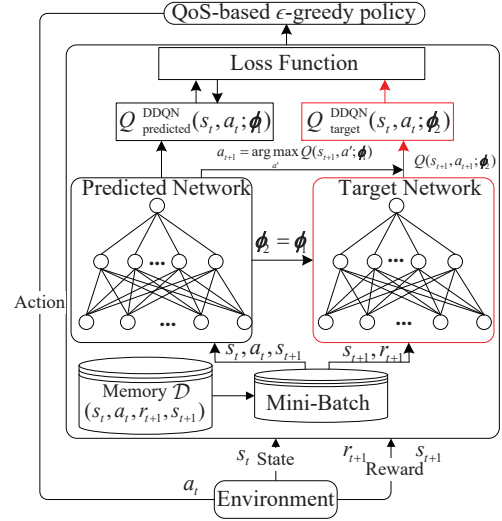}
\captionsetup{font={small}}
\caption{The DDQN structure with QoS-based $\epsilon$-greedy policy.}\label{Fig.R1 DQN}
\end{figure}

\section{Proposed Algorithm}
In this paper, we employ the RL algorithm to explore the unknown environment, where the UAV performs actions with the aim of maximizing the long-term system rewards by trying different actions, learning from the feedback, and then reinforcing the actions until the actions deliver the best result.
Furthermore, we use DDQN of DRL algorithm to address not only the overestimation problem of DQN, but also the massive state-action  pairs induced by time-varying TU locations  rather than conventional RL algorithm.
Besides, we develop a QoS-based $\epsilon$-greedy policy in our proposed algorithm to further meet the second constraint in (\ref{constraint}).

\subsection{Deep Q-Network (DQN)}
The state-action value function is $Q({s_t},{a_t}) = \mathbb{E}\left[ {\sum\nolimits_{\hat t = t}^{T-1} {\omega {r_{\hat t + 1}}|{s_{\hat t}},{a_{\hat t}}} } \right]$, where $\omega\in[0,1]$ is the discount factor and ${r_{\hat t + 1}}$ is the immediate reward in the $\hat t$th time slot based on the state-action pair (${s_{\hat t}},{a_{\hat t}}$) \cite{DDQN}.
The concept of $Q({s_t},{a_t})$ is to evaluate how good the action $a_t$ performed by the UAV in the state $s_t$ is.
 As illustrated in \cite{DDQN}, DQN approximates the Q-value by using two deep neural networks (DNNs) with the same four fully connected layers but different parameters $\bm{\phi_1}$ and $\bm{\phi_2}$.
One is the predicted network, whose input is the current state-action pair $(s_t,a_t)$ and output is the predicted value, i.e., $Q_{{\rm{predicted}}}^{{\rm{DQN}}}(s_t,a_t;\bm{\phi_1})$.
The other one is the target network, whose input is the next state $s_{t+1}$ and output is the maximum Q-value of the next state-action pair. Given this output, the target value of $(s_t,a_t)$ is $Q_{{\rm{target}}}^{{\rm{DQN}}}(s_t,a_t;\bm{\phi_2}) = {r_{t + 1}} + \omega \mathop {\max }\limits_{a'} Q({s_{t + 1}},a';{\bm{\phi _2}}) \label{DQN_target}$, where $a'$ is the candidate of next action.

\subsection{DDQN with Proposed QoS-Based $\epsilon$-greedy Policy}
DQN structure chooses $\mathop {\max }\limits_{a'} Q({s_{t + 1}},a';{\bm {\phi_2}})$ directly in the target network, whose parameter is not updated timely and may lead to the overestimation of Q-value \cite{DDQN}.
To address the overestimation problem, DDQN applies two independent estimators to approximate the Q-value.
 Fig. 2 shows the DDQN structure with QoS-based $\epsilon$-greedy policy.
The predicted network outputs $Q_{{\rm{predicted}}}^{{\rm{DDQN}}} (s_t,a_t;\bm \phi_1)$.
For the target network, DDQN chooses the action for the next state that yields $\mathop {\arg \text{max} }\limits_{a'} Q({s_{t + 1}},a';{\bm {\phi_1}})$ in the predicted network and identifies the corresponding Q-value of next state-action pair in  the target network, i.e., $Q({s_{t + 1}},\mathop {\arg \text{max} }\limits_{a'} Q({s_{t + 1}},a';{\bm {\phi_1}});{\bm {\phi_2}})$.
Consequently, the target value in DDQN is defined as
\begin{align}
&Q_{{\rm{target}}}^{{\rm{DDQN}}} (s_t,a_t;\bm \phi_2) \nonumber \\
&\!=\! {r_{t + 1}} \!+\! \omega Q({s_{t + 1}},\mathop {\arg \text{max} }\limits_{a'} Q({s_{t + 1}},a';{\bm {\phi_1}});{\bm {\phi_2}}) \label{DDQN_target}.
\end{align}
 The goal of the two DNNs is to approximate the Q-value in (12).
Based on this Q-value, the UAV chooses an action $a_t$ according to the current state $s_t$ with the proposed QoS-based $\epsilon$-greedy policy, receives the reward $r_{t+1}$, and then transfers to the next state $s_{t+1}$. At time slot $t$,  a transition pair is defined as $(s_t,a_t,r_{t + 1},s_{t + 1})$.

The description of the DDQN structure is given in Algorithm 1.
From lines 11 to 19, the  DNNs are trained by the transition pairs   stored in memory $\mathcal D$.
In line 12, $K$ mini-batch samples are randomly extracted from $\mathcal D$ to update ${\bm {\phi_1}}$.
In line 16, the loss function is $J({\bm {\phi_1}}) = \frac{1}{{2K}}\sum\nolimits_{k = 1}^K {{{[ {Q_{{\rm{target}}}^{{\rm{DDQN}}}(k) - {Q_{{\rm{predicted}}}^{{\rm{DDQN}}}}(k)} ]}^2}}$, where $Q_{{\rm{target}}}^{{\rm{DDQN}}}(k)$ and ${Q_{{\rm{predicted}}}^{{\rm{DDQN}}}}(k)$ represent the target and predicted values of the $k$th sample from the $K$ mini-batch samples, respectively.
In line 17, the gradient descent method is applied to update ${\bm {\phi_1}}$ of the predicted network as ${\bm {\phi_1}} = {\bm {\phi_1}} - \lambda {\nabla _{{\bm {\phi_1}}}}J({\bm {\phi_1}}),\label{PHI_1}$
where $\lambda\in[0,1]$ is the learning rate and $\nabla_{{\bm {\phi_1}}}$ is the gradient function with respect to $\bm {\phi_1}$.
Moreover, ${\bm {\phi_2}}$ is updated as ${\bm {\phi_2}} = {\bm {\phi_1}}$ after a fixed interval.
To achieve a good tradeoff between exploration and exploitation, a decrement $\delta$ is subtracted from $\epsilon$ in line 20.
The episode ends in the $T-1$th time slot if $b(T-1)\leq 0$. Finally, the proposed algorithm produces the optimal policy $\pi^*$ in line 24.

\begin{algorithm}[t]
\begin{small}
\caption{The DDQN with QoS-based $\epsilon$-greedy policy}
  \begin{algorithmic}[1]
    \State \textbf{Initialization:} $\bm{\phi_1}$ and $\bm{\phi_2}$; $\epsilon$, $\delta$, and $K$;
    \For{$j=0 \ \text{to}\ N_\text{e}$ ($N_\text{e}$ is the number of episodes)}
        \State Let $t=0$, $T=0$, and get the initial state $s_t$;
        \While{$b(t) > 0$}
            \State Take action $a_t$ with \underline{QoS-based $\epsilon$-greedy policy} at $s_t$;
            \State {\kern 1pt} {\kern 1pt} {\kern 1pt} Case I: {\kern 1pt} $a_t$ = $a_{nm}(t)$ with $\epsilon$-greedy policy;
            \State {\kern 1pt} {\kern 1pt} {\kern 1pt} Case II: If $n \in \mathcal{N}_{\text{nQoS}}$, $a_t = a_{nm}(t)$ ;
            \State {\kern 1pt} {\kern 1pt} {\kern 1pt} {\kern 1pt} {\kern 1pt} {\kern 1pt} {\kern 1pt} {\kern 1pt} {\kern 1pt} {\kern 1pt} {\kern 1pt} else $a_t = a_{n'm'}(t)$ until $n'\in \mathcal{N}_{\text{nQoS}}$;
            \State Obtain the reward $r_{t+1}$ and transfer to $s_{t + 1}$;
            \State Store the transition $(s_t,a_t,r_{t + 1},s_{t + 1})$ in the memory $\mathcal D$;
            \If {$\mathcal D$ is full}
                \State Randomly extract $K$ mini-batch samples from $\mathcal D$;
                \For {$k=1 {\kern 1pt}{\kern 1pt}{\kern 1pt}\text{to}{\kern 1pt}{\kern 1pt}{\kern 1pt} K$}
                \State Obtain ${Q_{{\rm{predicted}}}^{{\rm{DDQN}}}}(k)$ and $Q_{{\rm{target}}}^{{\rm{DDQN}}}(k)$;
                \EndFor
                \State Let $J({\bm{\phi _1}}) = \frac{1}{{2K}}\sum\limits_{k = 1}^{K}{{{[ {Q_{{\rm{target}}}^{{\rm{DDQN}}}(k) - {Q_{{\rm{predicted}}}^{{\rm{DDQN}}}}(k)} ]}^2}}$;
                \State Update ${\bm{\phi_1}}$ with ${\bm{\phi_1}}\leftarrow{\bm{\phi_1}} - \lambda {\nabla _{{\bm{\phi_1}}}}J({\bm{\phi_1}})$;
                \State After a fixed interval, update ${\bm{\phi_2}}$ as $\bm{\phi_2} = \bm{\phi_1}$;
            \EndIf
            \State Let $b(t+1)=b(t)-W(t)$, $t\leftarrow t + 1$, and $\epsilon \leftarrow \epsilon - \delta$;
        \EndWhile
        \State $T=t$;
    \EndFor
    \State \textbf{Output:} The optimal policy $\pi^*$.
  \end{algorithmic}
\end{small}
\end{algorithm}

\begin{figure}[ht]
\setlength{\abovecaptionskip}{0.cm}
\setlength{\belowcaptionskip}{-0.cm}
\centering
\includegraphics[width=2.8in,height=2.1in]{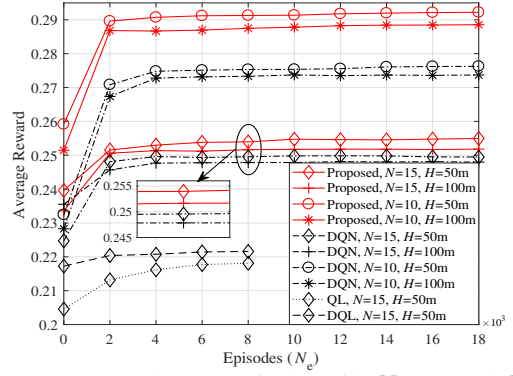}
\captionsetup{font={small}}         
\caption{Average reward versus episodes with $N=10$ and $N=15$.}\label{fig:average_reward}
\end{figure}

For the current state $s_t$, the UAV uses conventional $\epsilon$-greedy policy to select a random action $a_t$ with probability $\epsilon$ and ${a_t} = \mathop {\arg \text{max} }\limits_{a'} Q({s_t},a')$ with probability $1-\epsilon$, which is unable to guarantee the QoS constraint in (\ref{constraint}).
Consider an arbitrary time slot $t$. To meet the QoS constraint, we develop a QoS-based $\epsilon$-greedy policy to choose the optimal action of $s_{t}$ from lines 5 to 8 in Algorithm 1 as follows:

 Case I: $\sum\nolimits_{\hat{t} = 0}^{t-1} {{\mu _n}(\hat{t}) \ge Z}, \forall n \in \{ 1,...,N\}$
In this case, all TUs satisfy the QoS constraint. Then the UAV chooses an action $a_{nm}(t)$ with conventional $\epsilon$-greedy policy.

 Case II: $\sum\nolimits_{\hat{t} = 0}^{t-1} {{\mu _n}(\hat{t}) < Z}, \exists n \in \{ 1,...,N\}$
In this case, there exists at least one TU that does not meet the QoS constraint in the $t$th time slot.
    First, the UAV collects the TUs in $\mathcal{N}_{\text{nQoS}}=\{n_1, n_2,...,n_I\}$ with $\sum\nolimits_{\hat{t} = 0}^{t-1} {{\mu _{n{_i}}}(\hat{t}) < Z}, \forall n_i \in \mathcal{N}_{\text{nQoS}}$.
    Then, the UAV chooses an action $a_{nm}(t)$ with conventional $\epsilon$-greedy policy.
    The UAV chooses the action $a_{nm}(t)$ if the associated TU $n\in\mathcal{N}_{\text{nQoS}}$ based on this $a_{nm}(t)$.
    Otherwise, the UAV discards this action $a_{nm}(t)$ and chooses another action $a_{n'm'}(t)$ until the associated TU $n'\in \mathcal{N}_{\text{nQoS}}$.

Note that Algorithm 1 describes the offline training process to find the optimal policy $\pi^*$.
Then  $\pi^*$ is used to instruct the UAV to serve the TUs with the maximal long-term system reward during the online testing process.

\emph{Remark 1:}
First, the Q-learning used in \cite{1-1} is not well-suited to our complex environment with real-time mobile TUs, since the number of state-action pairs increases over time and the cost of managing the Q-table is unaffordable.
Second, \cite{1-2} employed the dueling DQN to optimize the UAV deployment in the multi-UAV wireless networks, while our work uses the double DQN (DDQN) to optimize the UAV trajectory in the UAV-mounted MEC networks.
Third, different from the DQN-based UAV navigation in \cite{1-3}, we employ the DDQN-based algorithm to address the overestimation problem. $\blacksquare$

\begin{figure}[t]
\centering
\includegraphics[width=2.8in,height=2.1in]{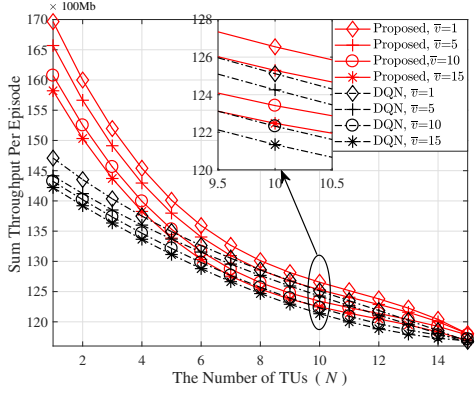}
\captionsetup{font={small}}
\caption{Sum throughput per episode versus the number of TUs with $H=50$m.}\label{fig:episode_throughut}
\end{figure}

\begin{figure}[t]
\centering
\includegraphics[width=2.8in,height=2.0in]{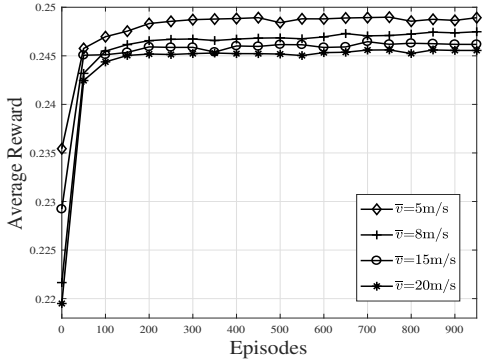}
\captionsetup{font={small}}
\caption{The robustness of our proposed algorithm under different average speeds of TUs with $H=50$m.}\label{Fig.1-R2 DQN}
\end{figure}

\section{Simulations and Results}
The simulation parameters are set as $M=25$ FPAPs, $B=200$\text {kJ}, $V=20$m/s, $\sigma^2=-140$\text{dB}, $\rho_0=-50$\text{dB}, $\gamma_\text c=10^{-27}$F, $C=1000$, $f_\text c=2$GHz, $N_\text b=100$Mb, $\epsilon=0.1$, $\delta=0.005$, $\eta=2$, $\beta=10$, $Z=5$, and $\mu_n(t)$ randomly ranges between 0 and 10 \cite{WQQ}.
The powers of each TU transmission, UAV flying and hovering are $P_\text{t}=0.1$W, $P_\text{f}=110$W and $P_\text{h}=80$W, respectively.

Fig. 3 depicts the average reward of proposed algorithm, DQN, DQL (double Q-learning), and QL algorithms.
First, our proposed algorithm achieves the largest convergence rate and average reward among all the algorithms.
Second, lower UAV altitude or less TUs contributes to a larger average reward.
On one hand, the higher UAV altitude results in larger path loss and more UAV hovering energy.
On the other hand, the UAV consumes more energy to meet the QoS constraint of each TU as the number of TUs goes up.
Third, when $N_\text{e}> 8000$, it is observed that QL and DQL are hardly implemented because the construction of the Q-table with massive states and actions is unaffordable.

Fig. 4 shows the sum throughput per episode of proposed algorithm and DQN algorithm versus the number of TUs.
We define the sum throughput per episode as the product of the offloaded tasks from all TUs per episode and the number of bits per task $N_\text{b}$.
First, the proposed algorithm achieves the largest sum throughput per episode among all the algorithms under any $\bar{v}$.
Second, the sum throughput per episode reduces as $N$ increases.
Third, the sum throughput per episode increases as $\bar{v}$ reduces for all the algorithms.
For example, both the proposed algorithm and DQN achieve their respective largest sum throughput per episode at $\bar{v}=1$m/s.
This is due to the factor that the path planning problem gradually reduces down to the problem with static TUs as $\bar{v}$ decreases, which can directly find the global optimal solution.

\begin{figure}[ht]
\centering
\includegraphics[height=0.65\columnwidth]{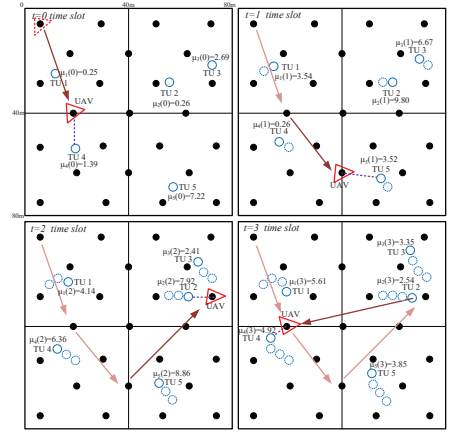}
\captionsetup{font={small}}
\caption{UAV trajectory design with $N=5$, $\bar{v}=1$m/s, and $H=50$m.}
\label{fig:trajectory}
\end{figure}

\begin{table}[ht]	
\center
\setlength{\tabcolsep}{2.8mm}\captionsetup{font={small}}
\caption{Percentage of QoS for $N=15$ TUs }
\subtable[Conventional $\epsilon$-greedy policy]{
\begin{tabular}{c|cccccccc}
\hline
TU index                    &1     &2      &3     &4     &5   \\  \hline
$\text {QoS} (\%)$ &16.296 &16.603 &22.871 &17.652 &17.759  \\ \hline
TU index                    &6     &7      &8      &9     &10   \\   \hline
$\text {QoS} (\%)$ &16.636 &15.742 &15.151 &12.768 &17.763 \\ \hline
TU index                   &11     &12    &13     &14    &15   \\  \hline
$\text {QoS} (\%)$ &13.794 &16.284 &16.286 &13.148 &13.783 \\  \hline
\end{tabular}}
\qquad
\subtable[ QoS-based $\epsilon$-greedy policy]{
\begin{tabular}{c|cccccccc}
\hline
TU index              &1     &2      &3     &4     &5   \\   \hline
$\text {QoS} (\%)$ &100.00 &99.999 &99.998 &99.997 &100.00  \\ \hline
TU index                 &6    &7      &8      &9     &10   \\   \hline
$\text {QoS} (\%)$ &99.998 &99.998 &99.998 &99.999 &99.996 \\ \hline
TU index                  &11     &12    &13     &14    &15   \\  \hline
$\text {QoS} (\%)$ &99.998 &99.998 &100.00 &99.999 &100.00 \\  \hline
\end{tabular}}
\end{table}

Fig. 5 shows that our proposed algorithm is robust under different average speeds of TUs.
Note that we only train the DNNs under $\bar{v}=1$ m/s and use the trained DNNs for $\bar{v}=5,8,15,20$m/s.
It is observed that the proposed algorithm can converge under speed variations.

Fig. 6 plots the UAV path planning with $N=5$ TUs and $\bar{v}=1$m/s from $t=0$ to $t=3$ based on the proposed algorithm.
The dashed and solid red triangles represent the initial and current locations of the UAV, respectively.
The black points are the projection of $M=25$ FPAPs.
The dashed and solid circles are the current and previous locations of each TU, respectively.
The dashed purple line links the UAV and its associated TU.
The arrows are the UAV trajectory.
It is shown that the UAV serves TU4 with $\mu_4(0)=1.39$, TU5 with $\mu_5(1)=3.52$, and TU2 with $\mu_2(2)=7.92$ in $t=0, 1, 2$ respectively.
To meet the QoS constraint with $Z=5$, the UAV flies back to serve TU4 with $\mu_4(3)=4.92$ in $t=3$.

Table I   presents the percentage of  QoS satisfaction over 100000 episodes for 15 TUs under conventional $\epsilon$-greedy policy and the proposed QoS-based $\epsilon$-greedy policy respectively.
It is observed that the proposed policy significantly outperforms conventional $\epsilon$-greedy policy.

\section{Conclusions}
We optimized the UAV trajectory in the UAV-mounted MEC network, where the UAV was deployed as a mobile edge server to dynamically serve the mobile TUs.
We formulated the optimization problem as an MDP, assuming that the motion of each TU follows the GMRM.
In particular, we developed the QoS-based $\epsilon$-greedy policy based on DDQN to maximize the long-term system reward and meet the QoS constraint.
The simulation results demonstrated that the proposed algorithm not only outperforms DQN, DQL and QL in terms of convergence and sum throughput, but also achieves almost $99\%$ guarantee rate in QoS of each TU.

\bibliographystyle{IEEEtran}
\bibliography{myreference}
\end{document}